\documentclass{amsart}

\usepackage[T1]{fontenc}
\usepackage[latin9]{inputenc}
\usepackage{graphicx}

\usepackage[table]{xcolor}

\usepackage{tikz}
\usetikzlibrary{arrows,automata}

\usepackage{tabularx}

\usepackage{sidecap}
\usepackage{booktabs}
\usepackage[nolist]{acronym}
\usepackage{xspace}

\definecolor{tablegray}{rgb}{0.85,0.85,0.85}

\newcommand{\ie}{i.\,e.\@\xspace}

\newcommand{\shalf}{{\textstyle\frac{1}{2}}}

\newcommand{\ddt}{\frac{d}{dt}}
\newcommand{\Arg}{\mathrm{Arg}}

\numberwithin{equation}{section}

\title[Synchronizability of networks with strongly delayed links]{Synchronizability of networks with strongly delayed links: a universal classification} 

\author{V. Flunkert}

\address{Institut f{\"u}r Theoretische Physik, TU Berlin,
  Hardenbergstra\ss{}e 36, 10623 Berlin, Germany}

\address{Instituto de Fisica Interdisciplinar y Sistemas
  Complejos,\\
  IFISC (UIB-CSIC), Campus Universitat de les Illes Balears, E-07122
  Palma de Mallorca, Spain}

\email{flunkert@itp.tu-berlin.de}

\author{S. Yanchuk}
\address{Institut f{\"u}r Mathematik, Humboldt
  Universit{\"a}t Berlin, Unter den Linden 6, 10099 Berlin, Germany}

\author{T. Dahms}
\address{Institut f{\"u}r Theoretische Physik,
  TU Berlin, Hardenbergstra\ss{}e 36, 10623 Berlin, Germany}

\author{E. Sch\"oll}
\address{Institut f{\"u}r Theoretische Physik,
  TU Berlin, Hardenbergstra\ss{}e 36, 10623 Berlin, Germany}

\begin{document}
\begin{abstract}
  We show that for large coupling delays the synchronizability of
  delay-coupled networks of identical units relates in a simple way to
  the spectral properties of the network topology.  The master
  stability function used to determine stability of synchronous
  solutions has a universal structure in the limit of large delay: it
  is rotationally symmetric around the origin and increases
  monotonically with the radius in the complex plane. We give details
  of the proof of this structure and discuss the resulting universal
  classification of networks with respect to their synchronization
  properties. We illustrate this classification by means of several
  prototype network topologies.
\end{abstract}
\maketitle
\tableofcontents

\section{Introduction}

Synchronization phenomena of coupled nonlinear oscillators are
omni\-present and play an important role in physical, chemical and
biological systems \cite{PIK01,BOC02,MOS02,BAL09}. Understanding the
synchronization mechanisms is crucial for many practical applications.

One of the most intriguing effects occurring in coupled nonlinear
systems is the synchronization of chaotic dynamics \cite{PEC90}. The
notions of chaos and synchronization apparently contradict each other.
A system is chaotic if small perturbations of the system's initial
condition are amplified resulting in an unpredictable dynamical
behavior.  Stable synchronization of two systems on the other hand
occurs when deviations between the system states decay with time
(negative transversal Lyapunov exponent).

The contradiction between these two characteristics is only apparent
because the decay and amplification occur in different directions in
phase space. Perturbations within the \ac{SM}, \ie, the manifold on
which the states of the two systems are identical, grow due to a
positive Lyapunov exponent within this manifold giving rise to the
chaotic dynamics. On the other hand, perturbations transversal to the
\ac{SM}, associated with deviations between the two systems, decay due
to a negative transverse Lyapunov exponent, thus leading to stable
synchronization. Another aspect is the influence of delayed coupling,
since time delay renders the phase space infinite-dimensional and may
on one hand induce instabilities and bifurcations, or may on the other
hand stabilize unstable states \cite{CHO07,SCH07,JUS09,ATA10}.

Semiconductor lasers are of particular interest in the study of chaos
synchronization. The synchronization properties may lead to potential
applications, e.g., to new secure communication schemes
\cite{FIS00,HEI02a,ARG05,KAN05,KIN08b,VIC07,KAN08a,KAN09,KIN10,ENG10}. However,
it is impossible to synchronize chaos in two delay-coupled systems
without self-feedback for large delays because the synchronized
solution is always transversely unstable.  In coupled lasers this
effect leads to spontaneous symmetry breaking, and only generalized
synchronization of leader-laggard type occurs \cite{MUL04}.

However, chaos synchronization of two delay coupled systems can be
stable if each system has self-feedback. For semiconductor lasers this
has been realized with a passive relay in form of a semitransparent
mirror or an active relay in form of a third laser in between the two
lasers \cite{FIS06,KLE06,SHA06,LAN07}. These structures are thus
interesting for chaos based communication systems.

For practical applications it is not only necessary that the
synchronized solution is (linearly) stable, but it is also important
how robust the synchronization is to noise. Here, nonlinear effects
may play an important role. In particular, for synchronized chaotic
systems bubbling \cite{ASH94,OTT94} plays a key role in this context.
This effect may lead to occasional desynchronization even for
arbitrarily small noise amplitudes.  Bubbling has been observed for
example in optical \cite{SAU98,TER99} and electrical \cite{GAU96}
systems. This paper reviews and extends our recent letter
\cite{FLU10b} on synchronization in delay-coupled networks with large
delay, providing a universal classification \cite{FLU11a}.

\section{Structure of the master stability function for large delay}

\label{sec:msf}
\let\orghdots\hdots
\let\orgddots\ddots
\let\orgvdots\vdots
\def\defaultscale{.5} 
\renewcommand*{\hdots}[1][\defaultscale]{$\scalebox{#1}{$\orghdots$}$}
\renewcommand*{\ddots}[1][\defaultscale]{$\scalebox{#1}{$\orgddots$}$}
\renewcommand*{\vdots}[1][\defaultscale]{$\scalebox{#1}{$\orgvdots$}$}

To determine the stability of a synchronized state in a network of
identical units, a powerful method has been developed
\cite{PEC98,PEC06} called the \ac{MSF}. The \ac{MSF} for networks with
coupling delays has been the subject of recent works
\cite{DHA04,KIN09,CHO09}. Time delay effects play an important role in
realistic networks. For example, the finite propagation time of light
between coupled semiconductor lasers
\cite{ERZ06a,CAR06,DHU08,FIS06,VIC08,WUE05a,FLU09,HIC11,LUE11b}
significantly influence the dynamics. Similar effects occur in
neuronal
\cite{ROS05,HAU07a,MAS08,SCH08,DAH08c,HOE09,BRA09,LEH11,HOE09a} and
biological \cite{TAK01} networks.

In this work we show \cite{FLU10b} that the \ac{MSF} for networks with
coupling delays has a very simple structure in the limit of large
delays. This allows us to prove a number of general statements about
the synchronizability of such networks.

We will first discuss the \ac{MSF} approach and since we are interested in
delay coupled systems, we will do this in the context of networks with
delay \cite{DHA04,KIN09,CHO09}.

\section{Stability of synchronized solutions}

Consider a system of $N$ identical units connected in a network with
a coupling delay $\tau$ \cite{KIN09}
\begin{equation}
  \ddt  x^i (t) =  f \left[ x^i(t)\right] + \sum_{j=1}^N g_{ij}  h\left[ x^j(t-\tau)\right] \label{eq:network}
\end{equation}
with $x^i\in\mathbb{R}^n$. Here, $g_{ij}\in\mathbb{R}$ is the coupling
matrix determining the coupling topology and the strength of each link
in the network, $f$ is the (non-linear) function describing the
dynamics of an isolated unit, and $h$ is a possibly non-linear
coupling function. A synchronized solution can only exist if the row
sum of the matrix is the same for each row, \ie, $\sigma =
\sum_{j=1}^N g_{ij}$ independent of $i$. In this case if the systems
start in a synchronized state $x^1=x^2=\dots=x^N=\overline x(t)$, they
will remain synchronized since the feedback term will be equal for all
$x^i$. The synchronized solution $\overline x(t)$ is then determined
by
\begin{equation}
  \ddt \overline x(t) = f\left[\overline x(t)\right] + \sigma h\left[\overline x(t-\tau)\right].
\end{equation}

To calculate the stability of this synchronized solution, we consider
the evolution of small perturbations $\xi^i(t)$
\begin{equation*}
  x^i(t) = \overline x(t) + \xi^i(t).
\end{equation*}
Inserting this ansatz into Eq.~\eqref{eq:network} and linearizing in
$\xi^i$ we find
\begin{equation}\label{eq:1}
  \ddt  \xi^i (t) =  Df \left[ \overline x(t)\right] \xi^i(t) + \sum_{j=1}^N g_{ij}  D h\left[ \overline x(t-\tau)\right] \xi^j(t-\tau),
\end{equation}
where $Df$ and $Dh$ are Jacobians.
Using the vector
\begin{equation*}
  \Xi(t) = (\xi^1(t),\, \xi^2(t),\dots,\, \xi^N(t))
\end{equation*}
the system of linear equations~(\ref{eq:1}) can be written as
\begin{equation}
  \ddt \Xi (t) = I_N\otimes Df\left[\overline x(t)\right] \Xi(t) + g \otimes Dh\left[\overline x(t-\tau)\right] \Xi(t-\tau), \label{eq:variation-network}
\end{equation}
where $I_N$ denotes the $N$-dimensional identity matrix and
$g=(g_{ij})$ is the coupling matrix.  We assume that the coupling
matrix $g$ is diagonalizable, i.e., there exists a unitary
transformation $U$ such that
\begin{equation*}
  U g U^{-1} = \mathrm{diag}(\sigma,\,\gamma_1,\, \gamma_2,\, \dots,\, \gamma_{N-1}).
\end{equation*}
Here, $\sigma$ is the row sum of $g$, which is always an eigenvalue of
$g$ corresponding to the eigenvector $(1,\, 1,\dots,\,1)$. We call
this the \textsl{longitudinal} eigenvalue of $g$. The other
eigenvalues $\gamma_k$ we then call the \textsl{transversal}
eigenvalues of $g$.

Diagonalizing $g$ in Eq.~\eqref{eq:variation-network} with the
transformation $U$ does not affect the term $I_N\otimes Df[\overline
x(t)]$ in Eq.~\eqref{eq:variation-network}, since this transformation
acts only on the matrix $I_N$. Thus after the diagonalization
Eq.~\eqref{eq:variation-network} is transformed into $N$ equations
\begin{align}
  \ddt \xi(t) &= D f\left[\overline x(t)\right] \xi(t) + \sigma\, D h\left[\overline x(t-\tau)\right] \xi(t-\tau), \label{eq:msf-var-sigma}\\
  \ddt \xi(t) &= D f\left[\overline x(t)\right] \xi(t) + \gamma_k\, D h\left[\overline x(t-\tau)\right] \xi(t-\tau) \label{eq:msf-var-gamma}
\end{align}
with $k=1,\dots,\,N-1$.  The first equation corresponds to
perturbations in the direction of the vector $(1,\, 1,\dots,\,1)$,
which act equally on each individual system and thus do not cause
desynchronization. A growing perturbation in this direction indicates
that the synchronized solution of the network is chaotic.

The $N-1$ other equations in~\eqref{eq:msf-var-gamma} describe
perturbations transversal to the \ac{SM}. The synchronized solution is
stable if and only if these perturbations decay, \ie, if the maximum
Lyapunov exponent arising from the variational
Eq.~\eqref{eq:msf-var-gamma} is negative for all transversal
eigenvalues $\gamma_k$.

Pecora and Carroll \cite{PEC90} defined a function $\lambda_{\rm max}:
\mathbb{C}\to\mathbb{R}$, which maps a complex number $re^{i\phi}$ to
the maximum Lyapunov exponent arising from the variational equation
\begin{align*}
  \ddt \xi(t) &= D f\left[\overline x(t)\right] \xi(t) + re^{i\phi}\, D h\left[\overline x(t-\tau)\right] \xi(t-\tau).
\end{align*}
This function is called the \acl{MSF} and it can be calculated
numerically. Once this function is known on a sufficiently large
domain in $\mathbb{C}$, one can immediately decide for any network
structure whether synchronization will be stable or not. One only
needs to evaluate the \ac{MSF} at the transversal eigenvalues
$\gamma_k$ of the particular network's coupling matrix. Thus the
condition for stable synchronization is then
\begin{equation}
  \label{eq:2}
  \lambda_{\max}(\gamma_k)<0, \quad k=1,\dots,N-1.
\end{equation}
This way the problem has been separated into one part that only
depends on the dynamics of the individual system, and another part
that only depends on the coupling topology.

We will now restrict our analysis to maps, but all ingredients of our
arguments are also valid for flows and we will point out where the
results differ slightly for flows.  Delay coupled maps have been
widely studied because they show similar behavior as delay
differential equation and interesting synchronization phenomena have
been found in these systems \cite{ATA04}.

For delay coupled maps the dynamics in the \ac{SM} is governed by the
equation
\begin{equation}
  \label{eq:3}
  x_{k+1} = f(x_k) + \sigma h(x_{k-\tau})  
\end{equation}
with $\tau\in\mathbb{N}$ and $x_k\in\mathbb{C}^d$ or $\in\mathbb{R}^d$
and the \ac{MSF} is calculated from
\begin{equation}
  \xi_{k+1} = Df(x_k) \xi_k + r e^{i\psi} Dh(x_{k-\tau}) \xi_{k-\tau} \label{eq:LD-MS}.
\end{equation}
Whether the synchronized dynamics is chaotic or not depends on whether
the \ac{MSF} evaluated at the eigenvalue $r e^{i\psi}=\sigma$, which
corresponds to perturbations parallel to the \ac{SM}, is positive or not.

With the matrix coefficients $A_k:=Df(x_k)$, and $B_k:=
Dh(x_{k-\tau})$  the variational equation is given by
\begin{equation}
  \xi_{k+1} = A_k \xi_k + r e^{i\psi}B_{k} \xi_{k-\tau}. \label{eq:LD-var}
\end{equation}

Note that when the delay is changed the dynamics in the \ac{SM} changes,
too.  Hence, we are not able to make predictions about what happens as
$\tau$ is changed. However, at a fixed large value of the delay time
$\tau$ we can compare the Lyapunov exponents arising from different
values of $re^{i\psi}$ in Eq.~\eqref{eq:LD-MS}.

We will now analyze the Lyapunov exponents arising from
Eq.~\eqref{eq:LD-var} in the limit of large $\tau$. We do this in the
following steps: first we analyse the two simpler cases, where the
dynamics in the \ac{SM} is a \ac{FP} or a \ac{PO}. Then to
expand the results to chaotic trajectories $x_k$ in the \ac{SM} we use
the fact that \acp{PO} are dense in a chaotic attractor.

\subsection{A fixed point in the synchronization manifold}
For \acp{FP} of delay differential equations there exists a scaling
theory for the \ac{FP}'s eigenvalues in the limit of large delay
\cite{GIA96,YAN05a,YAN06,WOL06,LIC11}. Recently this theory has been
generalized to the scaling of Floquet exponents \cite{YAN09}. In both
cases the eigenvalues or Floquet spectrum consist of two parts: a
strongly unstable part arising from unstable eigenvalues of the system
without delay and a pseudo-continuous spectrum, for which the real
part of the eigenvalues approach zero in the limit of large delay.
This scaling theory has been developed for flows. Since we restrict
ourselves to maps, we want to discuss the main ideas of the scaling
theory for maps now.  However, each step can be done in the same way
for flows by applying the large delay theory developed in
\cite{GIA96,YAN05a,YAN06,WOL06,YAN09}.

Let us first consider the case, where the dynamics in the \ac{SM} is a
\ac{FP}, \ie, a period $T=1$ orbit. In this case the coefficient
matrices in Eq.~\eqref{eq:LD-var} are constant $A=A_k$ and $B=B_k$.

Making the ansatz $\xi_k = z^k \xi_0$, we find an equation for the
multipliers $z$
\begin{equation}
  \chi(z)= \det[A -z I + re^{i\psi} B\, z^{-\tau}] = 0, \label{eq:charfp}
\end{equation}
where $I$ denotes the identity matrix.

For the strongly unstable spectrum
we suppose there is a solution with $|z|>1$. Then in the limit of
$\tau\to \infty$ Eq.~\eqref{eq:charfp} becomes
\begin{equation}
  \det[A -z I] = 0.
\end{equation}
Thus in the limit of large delay the eigenvalues $z$ of $A$ with
$|z|>1$ are also solutions of Eq.~\eqref{eq:charfp}.

We are now interested in the pseudo-continuous
spectrum, \ie, in the solutions with
$|z|\approx 1$ in the limit of large $\tau$.  We make the ansatz
$z=(1+\delta/\tau) e^{i\omega}$.  In the limit $\tau\to \infty$ we
have $(1+\delta/\tau)^{-\tau} \to e^{-\delta}$, and $(1+\delta/\tau)
\to 1$.  Thus in the limit $\tau\to\infty$ Eq.~\eqref{eq:charfp}
becomes
\begin{equation}
  0 = \det[A - I e^{i\omega} + r e^{-\delta} e^{i(\psi-\phi)} B ] \label{eq:LD-charlim}
\end{equation}
with $\phi=\omega\tau$. The complex equation~(\ref{eq:LD-charlim}) has
three variables $\phi$, $\omega$, and $\delta$. Hence, it determines
real-valued functions $\delta(\omega)$. As a result, it determines the
multipliers $z(\omega)$. Let us show how $\delta(\omega)$ can be
calculated. In particular, we will see that $\delta(\omega)$ is
independent on the phase $\psi$ of the variational equation.

If $B$ is invertible, we can calculate the eigenvalues
$\mu= r e^{-\delta} e^{i(\psi-\phi)}$ in the following equation
\begin{equation}
  0=\det[-B^{-1} (A-I e^{i\omega}) - \mu], \label{eq:invertible}
\end{equation}
which is a polynomial in $\mu$. This polynomial has exactly $d$ roots
$\mu_j$ ($j=1,\dots,\,d$), which are eigenvalues of $-B^{-1}(A-I
e^{i\omega})$. Since Eq.~(\ref{eq:invertible}) is independent of
$\psi$, $\mu$ does not depend on $\psi$ as well.

\begin{figure}[tb]
  \centering
  \includegraphics{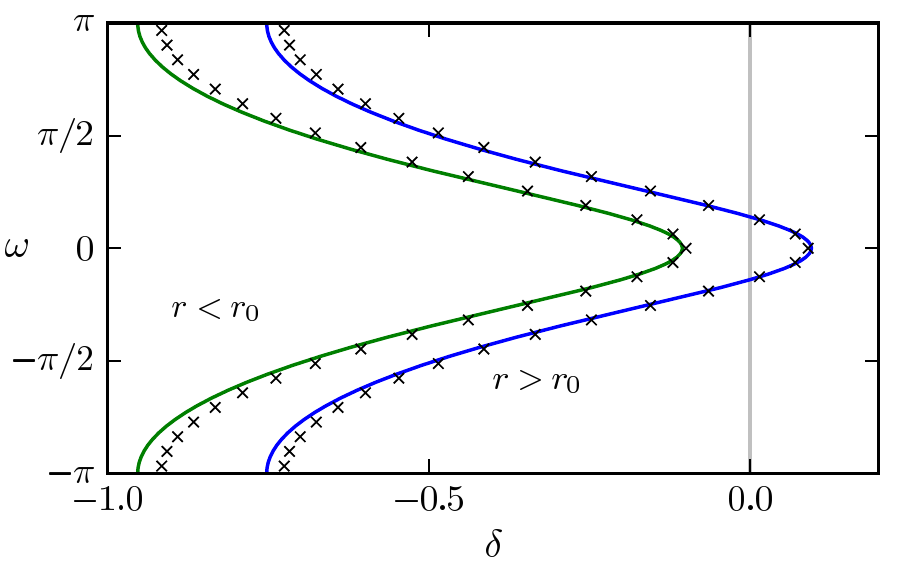}
  \caption[Pseudo continuous spectrum for a one dimensional
  map.]{\label{fig:pseudo} Pseudo continuous spectrum $\delta(\omega)$
    (lines) and location of the exact roots (crosses) for the example
    of a one dimensional complex map for $r=3.3 > r_0=3$ and $r = 2.7
    < r_0=3$.  Parameters: $a=0.4$, $b=0.2$, $\psi=0$, $\tau=30$.}
\end{figure}
If $B$ is not invertible, Eq.~\eqref{eq:LD-charlim} still gives a
polynomial in $\mu$, for which the roots can be calculated. Then each
eigenvalue $\mu$ will be a function of $\omega$ and one can find the
branches
\begin{equation*}
  \delta(\omega) = \ln \left[\frac{r}{|\mu(\omega)|}\right] = -\ln |\mu(\omega)| + \ln r.
\end{equation*}
The function $\mu(\omega)$ can admit the zero value at some point
$\omega_0$, \ie, $\mu(\omega_0)=0$, in the case when the matrix $A$
has an eigenvalue with $|z|=1$. Indeed, as follows from
Eq.~\eqref{eq:invertible}, for $\mu=0$, $\omega=\omega_0$ and $\det
B\ne 0$ we have
\begin{equation*}
  \det [ A- I e^{i\omega_0}] = \det[A-Iz]=0.
\end{equation*}
In all other cases, with $\det B\ne 0$ and $|z|\ne 1$, the function
$|\mu(\omega)|$ is bounded $0 < \mu_0 \le |\mu(\omega)| \le \mu_1$.

If there are no strongly unstable
eigenvalues, the sign of $\delta$
determines the stability in the limit of large $\tau$. It is clear,
that $\delta$ increases monotonically with increasing $r$ and in
particular $\delta$ is negative for small $r$ and positive for large
$r$. Thus there is a minimum radius $r_0$, for which the first
eigenvalue branch becomes unstable $\delta>0$ and thus the \ac{MSF}
changes sign.

Note that we have obtained the function $\delta(\omega)$ on which the
solutions lie in the limit of large $\tau$ but not yet the exact
values of $\omega$. These are not important in the limit of large
delay, since the eigenvalues become very dense on the curve
$\delta(\omega)$.  Indeed, the exact values of $\omega$ can be
calculated from the expression $\mu(\omega) = r e^{-\delta(\omega)}
e^{i(\psi - \omega\tau)}$, which implies
\begin{equation}
  \Arg\, \mu(\omega) = \psi -\omega\tau + 2\pi k \label{eq:arg}
\end{equation}
for any integer $k$.  Since $\mu(\omega)$ is a known eigenvalue of the
matrix $-B^{-1}(A-I e^{i\omega})$, Eq.~\eqref{eq:arg} can be
considered as a transcendental equation for determining the solutions
$\omega=\omega_k$.  In particular, Eq.~\eqref{eq:arg} implies that the
distance between the neighboring solutions $\omega_k$ and
$\omega_{k-1}$ reads
\begin{align*}
  \omega_k-\omega_{k-1} &= \frac{1}{\tau} \left[\Arg\, \mu(\omega_{k-1}) - \Arg\,\mu(\omega_k)\right] + \frac{2\pi}{\tau}\\
  &= 2\pi/\tau + \mathcal{O}\left(1/\tau^2\right).
\end{align*}
Thus it is proportional to $1/\tau$ and the curve $\delta(\omega)$ is
filled densely with roots as $\tau\to\infty$.

\begin{figure}[tb]
  \centering
  \includegraphics{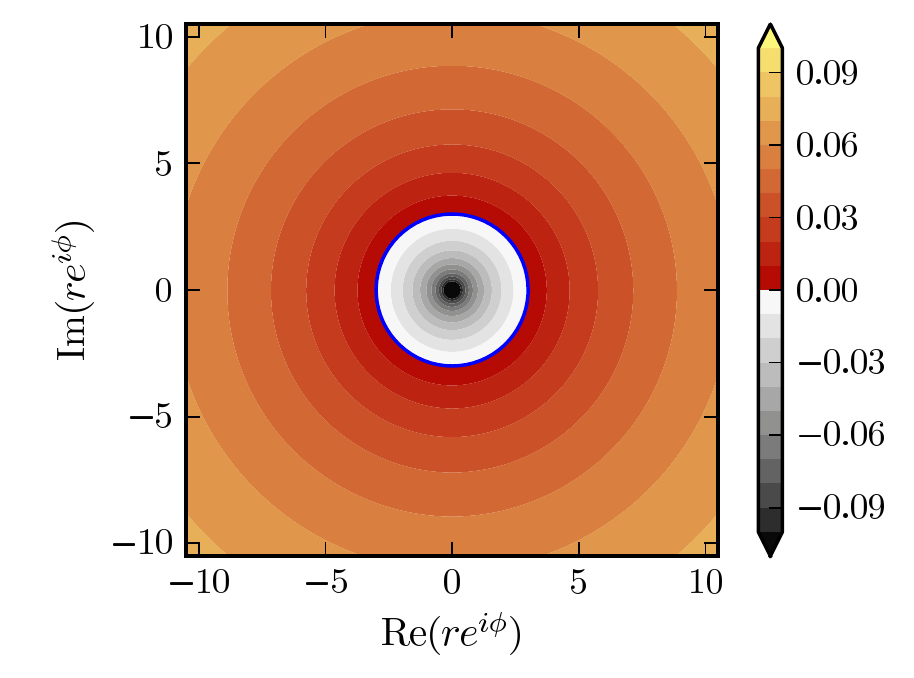}
  \caption[Master stability function for \ac{FP}.] {Master stability
    function for a one-dimensional map with \acl{FP} dynamics in the
    \acl{SM}. The red regions correspond to $\lambda_{\rm max}>0$
    (synchronized state is unstable). The gray regions correspond to
    $\lambda_{\rm max}<0$ (synchronized state is stable).  The blue
    circle indicates the stability boundary given by $r_0$ according
    to Eq.~\eqref{eq:LD-r0}. Already for relatively low values of
    $\tau$ the blue line matches very well the numerically obtained
    boundary.  Parameters of the variational equation: $a=0.4$,
    $b=0.2$, $\tau=20$.}
  \label{fig:msf-cartesian}
\end{figure}

Note that the curve $\delta(\omega)$ is determined in the bounded
interval $\omega\in[0,\,2\pi]$ in contrast to the case of delay
differential equations \cite{YAN05a}, where $\omega$ varies on the
whole axis $(-\infty, \, \infty)$.

\begin{figure}[tb]
  \centering
  \includegraphics{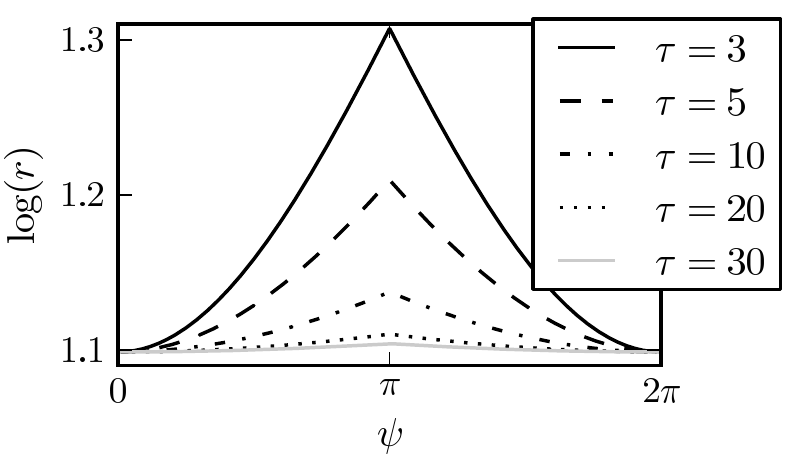}
  \caption[Critical radius $r(\psi)$ for different values of $\tau$.]
  { Boundary $r(\psi)$ of stability domain in polar coordinates for
    different values of $\tau$ in logarithmic scale. With increasing
    $\tau$ $r(\psi)\to r_0$.  Parameters of the variational equation:
    $a=0.4$, $b=0.2$.}
  \label{fig:msf-polar}
\end{figure}
The simple case is that of a one dimensional ($d=1$) complex map,
where $A=a\in\mathbb{C}$ and $B=b\in\mathbb{C}$ with $|a|<1$. In this
case we can explicitly calculate
\begin{equation*}
  \delta(\omega) = \ln(|r b|/|a-e^{i\omega}|).
\end{equation*}
For $r<(1-|a|)/|b|$ all the eigenvalues approach magnitude $1$ from
the stable side and for $r> (1-|a|)/|b|$ there are always weakly
unstable eigenvalues. Thus the \ac{MSF} changes sign at
\begin{equation}
  r_0=(1-|a|)/|b|. \label{eq:LD-r0}
\end{equation}
The pseudo-continuous spectrum for
these two cases is depicted in Fig.~\ref{fig:pseudo}.  The
corresponding \ac{MSF} is shown in Fig.~\ref{fig:msf-cartesian}. As
$\tau\to\infty$, the $\lambda_{\rm max}=0$ contour line approaches the
circle with radius $r_0$. This is depicted in
Fig.~\ref{fig:msf-polar}, where the angle-dependence of the critical
radius is shown for different values of $\tau$ in a logarithmic
scale. For small values of $\tau$, the critical radius has a strong
angle-dependency. However, already for a value of $\tau=20$, the
rotation symmetry is almost perfect (see
Fig.~\ref{fig:msf-cartesian}).

\subsection{A periodic orbit in the synchronization manifold}
Now consider the map
\begin{equation}
  \xi_{k+1} = A_k\, \xi_k + r e^{i\psi} B_k\, \xi_{k-\tau}, \label{eq:po}
\end{equation}
where $A_k$ and $B_k$ are periodic with period $T$, corresponding to a
\ac{PO} in the \ac{SM}. We consider the case of large delay
$\tau \gg T$.

Making a Floquet--like ansatz $\xi_k = z^k\, q_k$,
where $q_k$ is $T$ periodic we find
\begin{equation}
  z\, q_{k+1} = A_k\, q_k + re^{i\psi} B_k\, z^{-\tau} q_{k-n } \label{eq:floq}
\end{equation}
with $ n = \tau\, \mathrm{mod} \,T \in \lbrace 0,\,1,\dots,\,
T-1\rbrace$.

For the strongly unstable spectrum
again suppose there is a solution with $|z|>1$, then in the limit
$\tau\to\infty$ the term $z^{-\tau}$ vanishes and we find
\begin{equation}
  z\, q_{k+1} = A_k\,q_k. \label{eq:floq-strong}
\end{equation}
Using the periodicity of $q_k$, Eq.~\eqref{eq:floq-strong} implies
\begin{equation*}
  \det [ z^T - \prod_{k=1}^{T} A_k] = 0,
\end{equation*}
where $z^T$ is a Floquet multiplier of the system $\xi_{k+1} = A_k
\xi_k$ without delay.

Hence, if $z^T$ is a Floquet multiplier of Eq.~\eqref{eq:floq-strong},
with $|z|>1$, then in the limit $\tau\to\infty$ there exists also a
solution of Eq.~\eqref{eq:po}, which is close to $z$ and vice
versa. The rigorous proof of this fact as well as the convergence for
the pseudo-continuous spectrum is more tedious and we do not discuss
it here. The proof for the case of flows can be found in
Ref.~\cite{SIE11}.

For the pseudo-continuous spectrum
we again make the ansatz $z=(1+\delta/\tau) e^{i\omega}$ and taking
the limit $\tau\to\infty$ Eq.~\eqref{eq:floq} becomes
\begin{equation}
  e^{i\omega}\, q_{k+1} = A_k\, q_k + re^{-\delta} e^{i(\psi-\phi)} B_k\, q_{k - n}
\end{equation}
with $\phi=\omega\tau$.  One thus has to solve
\begin{equation}
  0 = [e^{i\omega} \overline J + \overline A + \mu \overline B]\vec q = 0, \label{eq:floq-pseudo}
\end{equation}
where $\overline A = \mathrm{diag}\lbrace A_1,\dots,\,A_T\rbrace$,
\newcommand{\pb}{\phantom{B_T}}
\begin{equation*}
  \overline J =
  \left[
    \begin{smallmatrix}
      0 &       & & I\\
      I &\ddots & & \\
      &\ddots &\ddots & \\
      &       & I & 0
    \end{smallmatrix}
  \right],
  \quad
  \overline B =
  \left[
    \begin{smallmatrix}
      0    & \pb   &        & B_1   &         & \\
      &\ddots & \pb    &      & \ddots  &\\
      &       & \ddots & \pb  &         & B_{n} \\
      B_{n+1} &       &        &\ddots& \pb     &\\
      &\ddots &        &      &\ddots   & \pb\\
      \pb   &       & B_T    &      &         & 0\\
    \end{smallmatrix}\right],
\end{equation*}
$\mu=re^{-\delta} e^{i(\psi-\phi)}$, and $\vec q = (q_1,\dots,\,q_T)$.
The position of the diagonal lines in the matrix $\overline B$ depends
on the value of $n=\tau\,\mathrm{mod} T$. Taking the determinant of
the matrix in Eq.~\eqref{eq:floq-pseudo} results in a polynomial in
$\mu=re^{-\delta}e^{i(\psi-\phi)}$ (of maximum order $d\times
T$). Again, the roots $\mu$ are functions of $\omega$ and we can
calculate the branches $\delta(\omega) = -\ln |\mu(\omega)| + \ln r$,
where $\psi$ and $\phi$ drop out. As in the case of \acp{FP}, one
can show that the function $|\mu(\omega)|$ is bounded $0<\mu_0\le
|\mu(\omega)| \le \mu_1$ unless the instantaneous system has a Floquet
multiplier $z$ with $|z|=1$.

We have again found the same structure of the \ac{MSF} : The \ac{MSF} is
rotationally symmetric in the complex plane about the origin. If
without feedback $(r=0)$ the \ac{MSF} is positive, then it is constant in
the limit of large delay. Otherwise it is a monotonically increasing
function of $r$ and there is a critical radius $r_0$ where it changes sign.

\subsection{A chaotic attractor in the synchronization manifold}
In every chaotic attractor there is an infinite number of \acp{UPO}
embedded. It has been long known that the characteristic properties of
the chaotic system can be described in terms of these \ac{PO}
\cite{CVI08}. Intuitively, the chaotic trajectory follows the
\acp{UPO} closely and ``switches'' between them, thus averaging over
the \acp{UPO} in the appropriate way allows us to calculate
statistical properties of the attractor.  One of the most important
examples is the natural measure of the chaotic attractor, which is
concentrated at the \ac{UPO} (hot-spots) and can in fact be expressed
in terms of the orbit's Floquet multipliers \cite{GRE88,LAI97}.

Lyapunov
exponents
arising from variational equations such as Eq.~\eqref{eq:LD-var} have
been discussed in the framework of periodic orbit
theory \cite{CVI93,CVI95,CVI08,ZAK10},
too.  In particular it has been shown \cite{NAG97} that a chaotic
attractor in an invariant manifold loses its transversal stability in
a blow-out bifurcation when the transversely unstable orbits outweigh
the transversely stable orbits. To be precise, we divide the orbits
into two groups of transversely stable and unstable orbits and define
\cite{NAG97} the transversely stable weight $\Lambda_T^{s}$ and the
unstable weight $\Lambda_T^{u}$ as
\begin{align}
  \Lambda_T^{u,s} &= \sum_{j=1}^{N^{u,s}_T} \mu_T(j) \lambda_T(j), \label{eq:weightedsum}
\end{align}
where the sum goes over all $N^u_T$ transversely unstable and $N^s_T$
transversely stable orbits of period $T$, respectively. Here,
$\mu_T(j)$ is the weight of the $j$th orbit, corresponding to the
natural measure of a typical trajectory in the neighborhood of the
$j$th orbit and $\lambda_T(j)$ is the transversal Lyapunov exponent of
this $j$th orbit. The weight of a \ac{PO} is inversely proportional to
the product of its unstable Floquet multipliers
\cite{GRE88,ZAK10}. The attractor is transversely unstable if and only
if in the limit of large $T$
\begin{equation}
  \Lambda^u_T  > |\Lambda^s_T|. \label{eq:weightcond}
\end{equation}

We now make the connection to the scaling theory for large $\tau$.
Starting from $r=0$ (no feedback) transversal Lyapunov exponents
$\lambda_T(j)$ of each orbit can only increase with increasing $r$, as
shown above. In particular for large enough $r$ the orbits become
transversely unstable: either they are already unstable for $r=0$ and
thus remain unstable or the pseudo-continuous
spectrum goes to zero and for large
$r$ it does so from the unstable side. Thus there exists a minimum
radius $r_0$, for which the condition \eqref{eq:weightcond} on the
weights is fulfilled.  Note that since we consider the limit
$\tau\to\infty$ we can evaluate Eq.~\eqref{eq:weightcond} at
arbitrarily large $T$, although it is a common result of \ac{PO}
theory that formulas such as Eq.~\eqref{eq:weightedsum} converge
quickly.

Thus in summary the \ac{MSF} has the same structure as for \acp{FP}
and \acp{PO} (the rotation symmetry follows from the rotation symmetry
of each $\lambda_T(j)$).

\section{Consequences for synchronization of networks}
Let us now discuss what the structure of the \ac{MSF} means for the
synchronizability of networks. We can categorize networks into three
types depending on the magnitude of the largest transversal
eigenvalue $\gamma_{\rm max}$ in relation to the magnitude of the row sum
$\sigma$: (A) the largest transversal eigenvalue is strictly smaller
than the magnitude of the row sum ($|\gamma_{\rm max}| < |\sigma|$),
(B) the largest transversal eigenvalue has the same magnitude as the
row sum ($|\gamma_{\rm max}| = |\sigma|$), and (C) the largest
transversal eigenvalue has a larger magnitude than the row sum
($|\gamma_{\rm max}| > |\sigma|$).

At $r=|\sigma|$ the \ac{MSF} is positive ($r_0 < |\sigma|$) for
chaotic dynamics in the \ac{SM} and negative ($|\sigma|<r_0$) for
dynamics on a stable \ac{PO} or a \ac{FP}. This gives us a lower or an
upper bound on $r_0$ and we can thus give the classification as shown
in Table \ref{table}.
\begin{table}
  \caption{\label{table} Stability of chaotic and non-chaotic synchronized
    solutions for the three types of networks.}
  \centering
  \renewcommand{\arraystretch}{1.2}
  \rowcolors{1}{white}{tablegray}
  \begin{tabular}{p{2.7cm} p{3.5cm} p{3.5cm}}
    \toprule
    & chaotic dynamics\newline in the \ac{SM} ($r_0< |\sigma|$) & \ac{PO} or \ac{FP} in the \ac{SM}\newline ($|\sigma| < r_0$) \\
    \midrule
    (A)~~ $ |\gamma_{\rm max}|<|\sigma|$ & synchr.\ stable if \newline $ |\gamma_{\rm max}| < r_0 $ & synchr.\ stable \\
    (B)~~ $|\gamma_{\rm max}|=|\sigma|$  & synchr.\ unstable  & synchr.\ stable \newline \\
    (C)~~ $|\gamma_{\rm max}|>|\sigma|$ & synchr.\ unstable & synchr.\ stable\newline if $|\gamma_{\rm max}| < r_0$ \\
    \bottomrule
  \end{tabular}
\end{table}
In networks of type (A) and (B) synchronization on a \ac{FP} or a
\ac{PO}, which is stable within the \ac{SM}, is always stable. For
type (C) this dynamics may be stable or not depending on the
particular network (value of $|\gamma_{\rm max}|$) and the dynamics in
the \ac{SM} (value of $r_0$). On the other hand chaos
synchronization is always unstable in
networks of type (B) and (C) and it may be stable or not in networks
of type (A) again depending on the particular network and the
dynamics.

Note that for autonomous flows with a stable \ac{PO} in the \ac{SM} we
always have $r_0=|\sigma|$, due to the \ac{PO}'s Goldstone
mode.  Thus for this case we cannot decide
whether synchronization for type (B) networks will be stable or
not. This depends on whether the $\lambda_{\rm max}=0$ contour line of
the \ac{MSF} approaches the circle with radius $r_0=|\sigma|$,
locally, at the transversal eigenvalues with $|\gamma_k|=|\sigma|$,
from the outside (stable) or from the inside (unstable).

\begin{figure}
  \centering
    \includegraphics[width=0.6\textwidth]{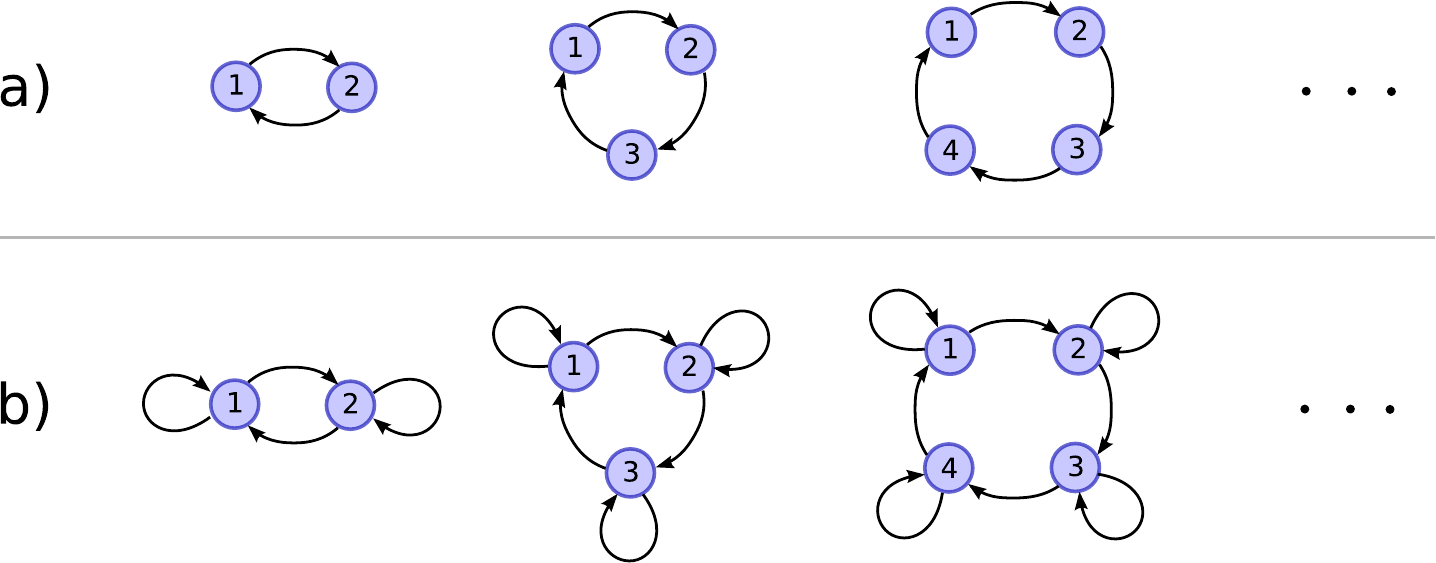}
    \caption{(a) unidirectionally coupled rings without feedback:
      always class (B) (no chaos synchronization), (b)
      unidirectionally coupled rings with feedback (for positive
      coupling): class (A) (chaos synchronization possible).}
  \label{fig:motifs}
\end{figure}
We now list some examples for the three types of networks (Fig.~\ref{fig:motifs}). The
categorization follows from the eigenvalue structure (spectral radius)
for the corresponding matrices, which can, for instance, be derived
using Gerschgorin's theorem.
\begin{itemize}
\item Mean field coupled systems (all-to-all coupling) have $\gamma_k
  = 0$ for all $k$ and are thus of type (A).
\item Networks with only inhibitory connections (negative entries)
  or only excitatory connections (positive entries) are up to the
  row sum factor stochastic matrices, \ie, the coupling matrix $G$ can
  be written as
  \begin{equation*}
    G = \sigma P \;,
  \end{equation*}
  where $P$ is a stochastic matrix (non-negative entries and unity row
  sum).  For stochastic matrices it is well known that the spectral
  radius is one, \ie, all eigenvalues have magnitude smaller than or
  equal to one. The proof utilizes Gerschgorin's theorem
  \cite{LEH10}. Thus it follows for $G$ that no eigenvalues has
  magnitude larger than $|\sigma|$ and these networks are of type (A)
  or (B).
\item Any network with zero row sum ($\sigma=0$) is of type (B)
  (trivial case) or (C).
\item Rings of uni-directionally coupled elements and two
  bidirectionally coupled elements are of type (B) \cite{CHO09}.
\item Unidirectionally coupled rings with self-feedback are for
  positive couplings of type (A).
\item Bidirectionally coupled rings with even number of elements
  without self-feedback are of type (B).
\item Bidirectionally coupled rings with odd number of elements are of
  type (A).
\end{itemize}
In the literature there is a great amount of material on the relation
of the spectral radius and the row sum for certain types of
matrices. These results are immediately applicable to our classification.
For a concrete network topology the classification is of course very
simple.

Networks with $\sigma=0$ belong to class (B) (trivial case) or to
class (C).  This confirms the conjecture stated in \cite{KIN09}:
Networks for which the trajectory of an uncoupled unit is also a
solution of the network ($\sigma=0$) cannot exhibit chaos
synchronization for large coupling delay.

For the chaotic case there may exist another radius $r_b$, with $0 \le
r_b\le r_0$, where the first \ac{PO} in the attractor loses its
transverse stability and the attractor undergoes a
bubbling bifurcation \cite{OTT94,ASH96a,FLU09}. Then
any network with $r_b < |\gamma_{\rm max}| < r_0$ will exhibit
bubbling in the presence of noise (or parameter mismatch), while any
network with $|\gamma_{\rm max}|<r_b$ will show stable
synchronization, even in the presence of noise.

\begin{figure}
  \centering
  \includegraphics[width=0.9\textwidth]{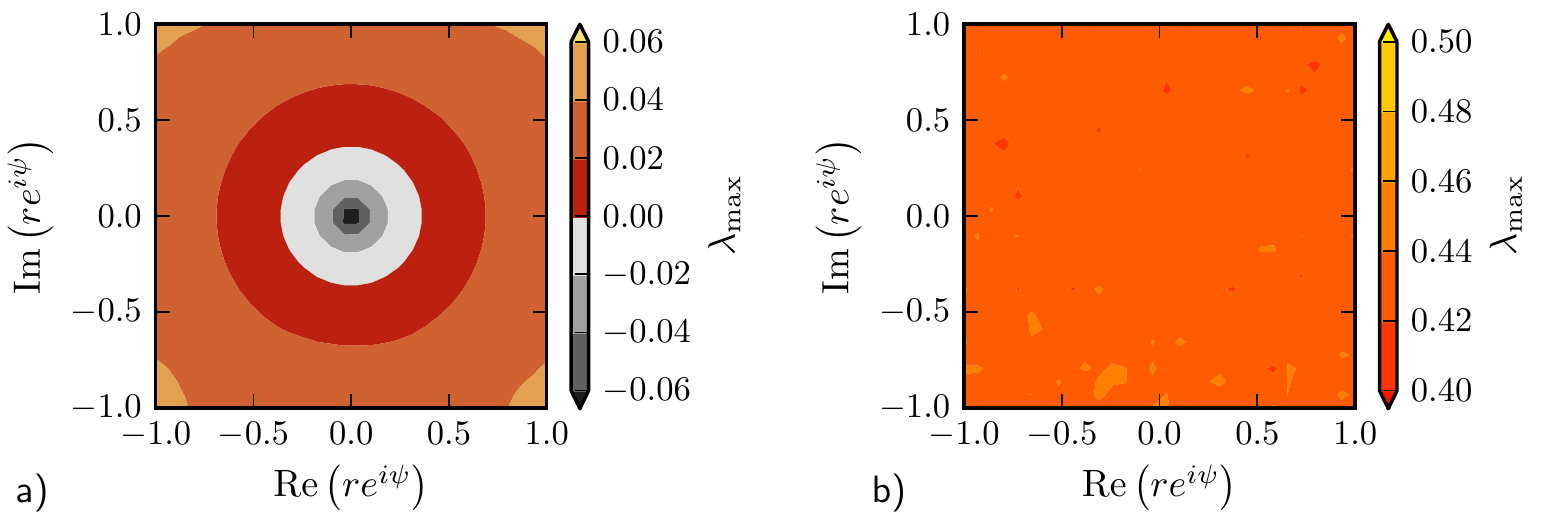}
  \caption[Master stability function for logistic maps.]{
    \label{fig:Master-stability-function}
    Master stability function (shown in color) for delay coupled
    logistic maps (Eq.~\eqref{eq:MSF}). Panel~(a) corresponds to
    $\lambda=3.2$ and the stable period-2 orbit within the
    \ac{SM}. Panel~(b) corresponds to $\lambda=3.8$ and a chaotic
    attractor within the \ac{SM}. In both cases the delay is chosen as
    $\tau=30$. }
\end{figure}

In order to illustrate the obtained results, let us consider the
following example of linearly coupled logistic maps
\begin{equation}
  x_{k+1}^{m}=\lambda
  x_{k}^{m}(1-x_{k}^{m})+\sum_{j=1}^{N}g_{mj}x_{k-\tau}^{j}
\end{equation}
with the zero row sums $\sigma=0$. The \ac{MSF} is calculated from the
following delayed system
\begin{equation}
  \xi_{k+1}=\lambda(1-2x_{k})\xi_{k}+re^{i\psi}\xi_{k-\tau},\label{eq:MSF}
\end{equation}
where the dynamics on the synchronization manifold $x_{k}$ is
determined by the map $x_{k+1}=\lambda x_{k}(1-x_{k})$.
Figure~\ref{fig:Master-stability-function} shows numerically computed
\ac{MSF}, i.e., the largest Lyapunov exponent of the system
(\ref{eq:MSF}) for two different cases: $\lambda=3.2$ and
$\lambda=3.8$, which correspond to the stable period-2 state and
chaos, respectively. The delay is set to $\tau=30$. In both cases, the
\ac{MSF} are radially symmetric. In the stable periodic case (panel
(a)), there exists a critical radius $r_{0}$ where the \ac{MSF}
changes sign, which determines the synchronizability of a given
coupled system.  In the chaotic case (panel (b)) the \ac{MSF} is close
to a positive constant, i.e., any coupling configuration will be
unstable. 

For the parameters used in panel (b) the dynamics exhibits so-called
{\em strong chaos} \cite{HEI11}. Strong chaos is characterized by a
Lyapunov exponent that stays constant with increasing delay time. As
recently shown \cite{HEI11} chaos synchronization is not possible in
the large delay limit for strong chaos. The other case of {\em weak
  chaos} occurs when the largest Lyapunov exponent scales as $1/\tau$
for $\tau\to\infty$. In this case chaos synchronization is possible
and the critical radius is determined by
\begin{equation} 
  r_0 = |\sigma| e^{-\lambda_m \tau} \;,
\end{equation}
where $\lambda_m$ is the maximum Lyapunov exponent of the system
\cite{HEI11}.


\section{Experimental setup for finding the critical radius}
We now propose an experimental method for determining the critical
radius $r_0$. Consider two elements coupled in
the following network motif
\begin{align*}
  x^1_{k+1} &= f(x^1_k) + \mu h(x^1_{k-\tau}) + \nu h(x^2_{k-\tau}),\\
  x^2_{k+1} &= f(x^2_k) + \mu h(x^2_{k-\tau}) + \nu h(x^1_{k-\tau}),
\end{align*}
where $\mu$ and $\nu$ are the self feedback strengths and the coupling
strengths, respectively.  Suppose we are able to change the
self-feedback strengths $\mu$ and the coupling strengths $\nu$, for
example by using gray filters in an optical experiment.

Let us choose
\begin{align*}
  \mu=\shalf(\sigma+r) \qquad{\rm and} \qquad \nu=\shalf(\sigma-r).
\end{align*}
Then the dynamics in the \ac{SM} is given by
\begin{equation*}
  x_{k+1} = f(x_k) + \sigma h(x_{k-\tau}),
\end{equation*}
while the variational equation transverse to the
\ac{SM} is given by
\begin{equation*}
  \xi_{k+1} = Df(x_k)\xi_k + r Dh(x_{k-\tau})\xi_{k-\tau}.
\end{equation*}
Thus by changing $r$ (for fixed $\sigma$) and checking whether the two
elements synchronize we are able to probe the \ac{MSF} along the real
axis at the radius $r$. Due to the monotonicity we can use a
root-finding algorithm such as the bisection method to find $r_0$ to
high accuracy with little iterations of the experiment and without
knowledge of the functions $f$ and $h$.  We can repeat this procedure
for other values of $\sigma$ and obtain the critical radius as a
function of $r_0(\sigma)$.  Thus from this rather simple setup we can
decide for any network of these elements whether synchronization is
stable or not.

\begin{figure}[tbp]
  \centering
  \includegraphics[width=0.3\textwidth]{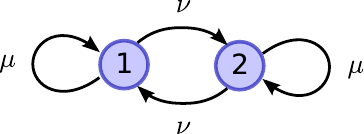}
  \caption{Schematic setup for determining the critical radius in an
    experiment.\label{fig:exp-r0}}
\end{figure}

As an example we consider two optoelectronically coupled
lasers
\begin{subequations}
  \label{eq:optoelectronic-coupling}
  \begin{align}
    \ddt \rho_1 &= n_1 \rho_1,\nonumber\\
    T \ddt n_1 &= p + \mu \rho_1(t-\tau) + \nu \rho_2(t-\tau) - n_1 - (1 + n_1) \rho_1,\\
    \ddt \rho_2 &= n_2 \rho_2,\nonumber\\
    T \ddt n_2 &= p + \mu \rho_2(t-\tau) + \nu \rho_1(t-\tau) - n_2 - (1 + n_2) \rho_2,
  \end{align}
\end{subequations}
where $\rho_i$ and $n_i$ is the intensity and the carrier density of
the $i$th laser, respectively.  The pump current of each laser is
modulated by the delayed intensities according to the coupling scheme
depicted in Fig.~\ref{fig:exp-r0}. Such feedback can be realized by
using photodiodes to measure the intensities of the arriving signals
and modulating the pump current accordingly. The bidirectional
coupling has strength $\nu$ and the self-feedback of each laser has
strength $\mu$.
For this setup we have numerically calculated $r_0(\sigma)$ in the
same manner as it would be done in an experiment: We choose a value of
$\sigma$, an interval $I_r = [r_{\rm min},\, r_{\rm max}]$ for the
$r$-domain and an initial value of $r=r_0$. We consider the systems to
be synchronized, if the relative synchronization error
\begin{equation*}
  \varepsilon:= \frac{\langle|\rho_1-\rho_2|\rangle}{\shalf[\langle \rho_1\rangle + \langle \rho_2 \rangle]}
\end{equation*}
is smaller than a threshold $\varepsilon_0$. We then simulate the system
and use the bisection method to find the synchronization threshold $r_0$
(up to a desired accuracy) in the interval $I_r$. We can then use the
calculated value of $r_0$ as an initial guess for neighboring
$\sigma$-values and thus follow the curve $r_0(\sigma)$.

\begin{figure}[tb]
  \centering
  \includegraphics{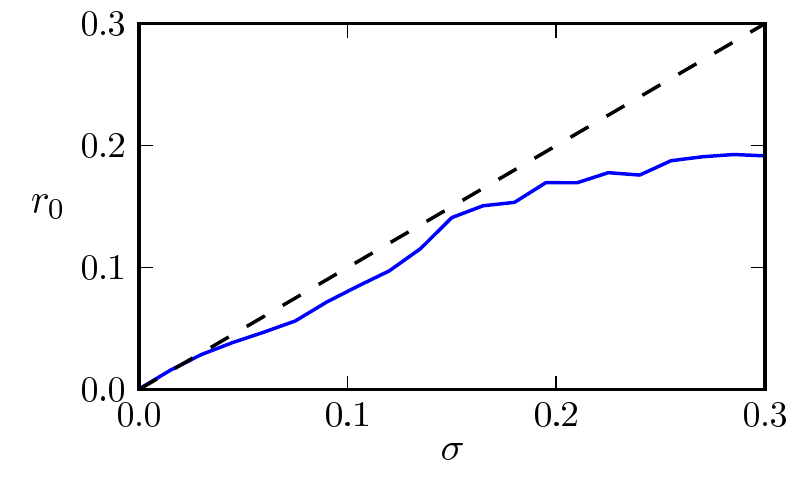}
  \caption[Critical radius $r_0$ as a function of $\sigma$.] {
    Numerically calculated critical radius $r_0$ as a function of
    $\sigma$ (solid curve) for the system of optoelectronically
    coupled lasers corresponding to
    Eqs.~\eqref{eq:optoelectronic-coupling}. A network can only have a
    stable synchronized solution if the magnitudes of its transversal
    eigenvalues are below the curve.  The curve is calculated up to an
    absolute error of $10^{-4}$. The dashed line shows the
    diagonal line $r_0=\sigma$.  \\
    Parameters: $\varepsilon_0=10^{-7}$, $p=1$, $T=200$,
    $\tau=2000$. }
  \label{fig:optoelek-r0}
\end{figure}

The result is depicted in Fig.~\ref{fig:optoelek-r0}, where the solid
curve shows $r_0(\sigma)$ and the dashed line corresponds to
$r_0=\sigma$. For small values of $\sigma$, \ie, weak feedback, the
dynamics is a \ac{PO} and due to the Goldstone mode we have $r_0
\approx \sigma$. For larger values of $\sigma$, the system becomes
chaotic and $r_0<\sigma$. For a given value of $\sigma$, a network has
a stable synchronized solution if and only if all transversal
eigenvalues $\gamma_k$ of the corresponding coupling matrix have
magnitude $|\gamma_k|<r_0(\sigma)$.

A similar method has been implemented in \cite{ILL11}. Here the structure of 
the \ac{MSF} was confirmed experimentally using optoelectronic oscillators.

\section{Conclusion and outlook}
In conclusion we have shown that the \ac{MSF} has a simple structure
in the limit of large delay: it is rotationally symmetric around the
origin and either positive and constant (if it is positive at the
origin) or monotonically increasing and becomes positive at a minimum
radius $r_0$. This structure allowed us to prove a recent conjecture
\cite{KIN09} about synchronizability of chaotic elements. Furthermore,
we classified network structures into three types depending on the
magnitude of the maximum transversal eigenvalue in relation to the
magnitude of the row sum and showed that these network types have
distinct synchronization properties. Using several prototype networks
like all-to-all or ring topologies we illustrated the scope of these
three classes. By means of a motif of two coupled nodes with feedback
loops, we proposed a very simple scheme with which the critical radius
$r_0$ can be found experimentally. Instead of mapping out the entire
domain of the master stability function, the knowledge of the
rotational symmetry allows the two-node motif to predict stability of
any network using the same local dynamics in the case of large delay
times.

The rotational symmetry of the \ac{MSF} has previously been found
numerically \cite{CHO09,LEH11}. In Ref.~\cite{LEH11} the same
structure of the \ac{MSF} has been found for a \ac{PO} in the \ac{SM}
for which the period $T$ is approximately equal to the delay time
$\tau$. For this case the structure of the \ac{MSF} has also been
derived analytically in \cite{LEH11}. Note that this case is
complementary to the situation $T\ll \tau$ that we looked at in this
paper. So the structure of the \ac{MSF} that we found seems to be
valid in even more general cases.

VF acknowledges financial support from the German Academic Exchange
Service (DAAD). This work was performed in the framework of SFB 910.


\begin{acronym}[MSF]
  \acro{MSF}{master stability function}
  \acro{SM}{synchronization manifold}
  \acro{UPO}{unstable periodic orbit}
  \acro{PO}{periodic orbit}
  \acro{FP}{fixed point}
\end{acronym}

\end{document}